\documentstyle[pre,aps,preprint]{revtex}

\begin{document} 

\draft 

\tighten

\preprint{\vbox{\hfill chao-dyn/9908009 \\
          \vbox{\hfill August 1999} \\
          \vbox{\hfill Revised December 1999} \\
          \vbox{\vskip0.5in}
         }}

\title{A Trace Formula for Products of Diagonal Matrix Elements
       in Chaotic Systems}

\author{Sanjay Hortikar\footnote{E--mail: \tt horti@physics.ucsb.edu}
        and
        Mark Srednicki\footnote{E--mail: \tt mark@physics.ucsb.edu} }

\address{Department of Physics, University of California,
         Santa Barbara, CA 93106 
         \\ \vskip1.0in          }

\maketitle

\begin{abstract}
\normalsize{
We derive a trace formula for $\sum_n A_{nn}B_{nn}\ldots\delta(E-E_n)$,
where $A_{nn}$ is the diagonal matrix element of the operator
$A$ in the energy basis of a chaotic system.
The result takes the form of a smooth term plus periodic-orbit corrections;
each orbit is weighted by the usual Gutzwiller factor times $A_p B_p \ldots$,
where $A_p$ is the average of the classical observable $A$ along the 
periodic orbit $p$.  This structure for the orbit corrections was previously
proposed by Main and Wunner on the basis of numerical evidence.  
}
\end{abstract}


\section{Introduction}

In a recent paper \cite{mw}, Main and Wunner introduced
the weighted density of states 
\begin{equation}
\rho^{(A,B,\ldots)} \equiv \sum_n A_{nn}B_{nn}\ldots\delta(E-E_n) \;.
\label{1}
\end{equation}
Here $A, B, \ldots$ are operators with smooth classical limits (Weyl symbols),
and $A_{nn}=\langle n|A|n\rangle$ is the diagonal matrix element of $A$ in 
the energy basis.  This is a simple generalization of 
$\rho^{(A)} \equiv \sum_n A_{nn}\delta(E-E_n)$, which has been studied 
extensively \cite{gutz,wilk,wilk2,efmw,em,efkamm,cr,crr}.
For chaotic systems, Main and Wunner proposed that
\begin{equation}
\rho^{(A,B,\ldots)} = \rho^{(A,B,\ldots)}_0 +
  {1\over\pi\hbar} 
  \sum_p A_p B_p \ldots w_p \;,
\label{2}
\end{equation}
where the sum is over all primitive periodic orbits $p$ with energy $E$, and
\begin{equation}
A_p \equiv {1\over \tau_p} \int_0^{\tau_p} A({\bf X}_p(t)) dt 
\label{ap}
\end{equation}
is the average value of the Weyl symbol of 
$A$ along the orbit; here $\tau_p$ is the period of the orbit,
and ${\bf X}=({\bf q},{\bf p})$ denotes both coordinates and momenta.  Also,
\begin{equation}
w_p \equiv \mathop{\rm Re} \sum_{r=1}^\infty 
                             {\tau_p e^{i(S_p/\hbar-\mu_p\pi/2)r}
                              \over |\det(M_p^r-I)|^{1/2}          } 
\label{w}
\end{equation}
is the Gutzwiller weight factor; $S_p$, $\mu_p$, and $M_p$ are 
respectively the action, Maslov index, and monodromy matrix 
of the orbit.  The first term on the right-hand side of Eq.~(\ref{2})
represents the part that remains smooth in the semiclassical limit;
it should be $O(\hbar^{-f})$, where $f$ is the number of freedoms.
However, Main and Wunner do not give an explicit formula for it.

If we set $B=\ldots=I$ in Eq.~(\ref{1}), and hence $B_p=\ldots=1$ in 
Eq.~(\ref{2}), then we recover the trace formula for 
$\rho^{(A)}$ \cite{wilk2,efmw,em,efkamm,cr,crr}.
If we set $A=I$ as well, we recover
the original Gutzwiller trace formula for the density of states \cite{gutz}.
This is the essential motivation of Main and Wunner for Eq.~(\ref{2}).
They provide strong numerical evidence in favor of it, but they do
not give an analytic derivation.  

In this paper, we remedy this situation by deriving Eq.~(\ref{2})
from a generalization of a trace formula originally due to 
Wilkinson \cite{wilk} (see also \cite{efmw,cr}).  
Furthermore we provide an explicit expression for the smooth term.

\section{Analysis}
\label{analysis}

We first consider the case of two operators, $A$ and $B$, and extend
the results to an arbitrary number in Section \ref{ext}.
Following Wilkinson \cite{wilk}, we define
\begin{equation}
S(E,\Delta) \equiv 
        \sum_{nm} A_{nm}B_{mn}\delta_1(E-{\textstyle{1\over2}}(E_n+E_m))
                           \delta_2(\Delta - (E_n-E_m)) \; .
\label{s1} 
\end{equation}
Here $\delta_1(E)$ and $\delta_2(E)$ are smeared delta-functions.
Rigorous results concerning $S(E,\Delta)$ have been proven in the case
that the Fourier transforms of these smeared delta functions have
compact support \cite{cr,crr}.  We will therefore make the simple choice
\begin{eqnarray}
\delta_i(E) &\equiv& \int_{-\tau_i}^{+\tau_i}{dt\over 2\pi\hbar}
                                         \,e^{iEt/\hbar} 
\label{delta1} \\ 
            &=& {\sin(E\tau_i/\hbar) \over \pi E} \;,
\label{delta2} 
\end{eqnarray}
where $\tau_i$, $i=1,2$, is a time cutoff.  Our results will come
from various manipulations of $S(E,\Delta)$ with $\Delta=0$.  

We begin by writing
\begin{eqnarray}
S(E,0) &=& \delta_2(0) \sum_n A_{nn}B_{nn}\delta_1(E_n-E)
\nonumber \\
     &&  {} + {1\over\pi}\sum_{n,m\ne n} A_{nm}B_{mn}\,
                   {\sin(\omega_{nm}\tau_2) \over E_n-E_m}\,
                   \delta_1(E-{\textstyle{1\over2}}(E_n+E_m)) \;,
\label{s2} 
\end{eqnarray}
where $\omega_{nm} \equiv (E_n-E_m)/\hbar$.  The first term on the
right-hand side is the one we want; except for the factor of
$\delta_2(0)=\tau_2/\pi\hbar$, it is the same as the right-hand
side of Eq.~(\ref{1}), in the limit as $\tau_1\to\infty$.  
To get rid of the unwanted second term, we take
$\tau_2$ to be much greater than the Heisenberg time 
$\tau_H \equiv 2\pi\hbar\rho_0$; here
\begin{equation}
\rho_0 \equiv \int {d^{2f}\!X\over(2\pi\hbar)^f} \, \delta(E-H({\bf X})) 
\label{rho0}
\end{equation}
is the Weyl formula for the mean density of states.  
If $\tau_2\gg\tau_H$, then 
we typically have $|\omega_{nm}|\tau_2 \gg 1$.  In this case, 
$\sin(\omega_{nm}\tau_2)$ varies erratically as $n$ and $m$ are varied.
Furthermore the factor of $1/(E_n-E_m)$ can be written as
$\rho_0/(n-m) = (2\pi\hbar/\tau_H)/(n-m)$, up to a factor
which also varies erratically.  We then have
\begin{eqnarray}
S(E,0) &=& {\tau_2\over\pi\hbar} \sum_{n} A_{nn}B_{nn}\delta_1(E_n-E)
\nonumber \\
     &&  {} + {\tau_H\over2\pi^2\hbar}\sum_{n,m\ne n} 
                   { A_{nm}B_{mn} R_{nm} \over n-m } \,
                   \delta_1(E-{\textstyle{1\over2}}(E_n+E_m)) \;,
\label{s3} 
\end{eqnarray}
where we can think of $R_{nm}$ as a random number.
Provided that $|A_{mn}|$ and $|B_{mn}|$ do not tend to increase
as $|m-n|$ increases (in general a decrease is to be expected), 
the sum in the second term should quickly converge.  Then we have
\begin{equation}
{\pi\hbar \over \tau_2}\, S(E,0) 
 = \sum_{n} A_{nn}B_{nn}\delta_1(E_n-E) + O(\tau_H/\tau_2) \;.
\label{s4} 
\end{equation}
The first term on the right-hand side is the same as the
right-hand side of Eq.~(\ref{1}), provided $\tau_1 \gg \tau_H$,
and the second term is small if $\tau_2 \gg \tau_H$.

We now wish to evaluate $S(E,\Delta)$ semiclassically.
We first use Eq.~(\ref{delta1}) in Eq.~({\ref{s1}) to get
\begin{equation}
S(E,\Delta) = 
\int_{-\tau_2}^{+\tau_2} {dt\over2\pi\hbar} \, 
                               e^{-i\Delta t /\hbar}
\int_{-\tau_1}^{+\tau_1} {dt'\over2\pi\hbar} \, 
                               e^{+iE t'\! /\hbar}
       \, F(t,t') \; ,
\label{sf} 
\end{equation}
where we have defined
\begin{equation}
F(t,t') \equiv \sum_{nm} A_{nm}B_{mn} \, e^{-i(E_n+E_m)t'\!/2\hbar} \, 
                                         e^{+i(E_n-E_m)t/\hbar} \;.
\label{f} 
\end{equation}
The key point is that we can write $F(t,t')$ as a single trace,
\begin{equation}
F(t,t') = \mathop{\rm Tr}U(-t+{\textstyle{1\over2}}t') A 
                         U( t+{\textstyle{1\over2}}t') B \;,
\label{tr} 
\end{equation}
where $U(t) = e^{-iHt/\hbar}$ is the time-evolution operator.

To simplify our exposition, we temporarily make the (otherwise unnecessary)
assumption that the Weyl symbols
of $A$ and $B$ are functions of only the coordinates
${\bf q}$ and not the momenta $\bf p$.  
We can then evaluate the trace by inserting two complete sets of
position eigenstates, leading to
\begin{equation}
F(t,t') = \int d^f \! q_1 \, d^f \! q_2 \,
          \langle {\bf q}_1| U(-t+{\textstyle{1\over2}t'}) |{\bf q}_2\rangle
          \, A({\bf q}_2) \,
          \langle {\bf q}_2| U( t+{\textstyle{1\over2}t'}) |{\bf q}_1\rangle
          \, B({\bf q}_1) \;.
\label{f2} 
\end{equation}
We now make use of the semiclassical approximation \cite{gutz,bm} to get 
\begin{eqnarray}
&& \int d^f \! q_2 \; 
       \langle {\bf q}_3| U(-t+{\textstyle{1\over2}t'}) |{\bf q}_2\rangle
       \, A({\bf q}_2) \,
       \langle {\bf q}_2| U( t+{\textstyle{1\over2}t'}) |{\bf q}_1\rangle
\nonumber \\
&& \qquad {} \cong 
       \sum_{\rm paths} K_{\rm path}({\bf q}_3,{\bf q}_1;t')
       \, A({\bf q}_{\rm path}(t+{\textstyle{1\over2}}t')) \; .
\label{bm}
\end{eqnarray}
Here the sum is over all classical paths that go from ${\bf q}_1$ at 
time zero to ${\bf q}_3$ at time $t'$, 
${\bf q}_{\rm path}(\tau)$ is the position reached
at time $\tau$ along a particular path,
and $K_{\rm path}({\bf q}_3,{\bf q}_1;t')$ is the contribution
of that path to the propagator 
$\langle {\bf q}_3| U(t') |{\bf q}_1\rangle$ 
in the semiclassical limit.

We now perform the integrals over $d^f \! q_1$ in Eq.~(\ref{f2}) and
over $dt'$ in Eq.~(\ref{sf}) by stationary 
phase \cite{gutz,wilk,wilk2,efmw,em,efkamm,cr,crr}.
We get a contribution from zero-length paths (for which $t'=0$
at the point of stationary phase), and a sum over contributions
from periodic orbits (for which $t'= \tau_p$ at the point of 
stationary phase).  The result is
\begin{equation}
S(E,\Delta) = \int_{-\tau_2}^{+\tau_2} {dt\over2\pi\hbar} 
                            \,  e^{-i\Delta t /\hbar}
              \left[ \rho_0 C_0(t)
                     + {1\over\pi\hbar} \sum_{\tau_p<\tau_1} 
                       w_p C_p(t) \right]\;,
\label{s5}
\end{equation}
where the sum is over all primitive periodic orbits with period
less than $\tau_1$.  Also, 
we have introduced the energy-surface correlation function
\begin{equation}
C_0(t) \equiv {1\over \rho_0} 
              \int {d^{2f}\!X\over(2\pi\hbar)^f} \,
              \delta(E-H({\bf X})) \, 
              A({\bf X}(t)) B({\bf X})    \;,
\label{c0}
\end{equation}
and the orbit correlation function
\begin{equation}
C_p(t) \equiv {1\over \tau_p} \int_0^{\tau_p} d\tau \, 
               A({\bf X}_p(\tau+t+{\textstyle{1\over2}}\tau_p)) 
               B({\bf X}_p(\tau)) \;.
\label{cp}
\end{equation}

Next, we must separate out a possible constant term in $C_0(t)$.
To do so, we take the microcanonical average of $A({\bf X})$ on
a surface of constant energy $E$,
\begin{equation}
A_0 \equiv {1\over \rho_0} 
              \int {d^{2f}\!X\over(2\pi\hbar)^f} \,
              \delta(E-H({\bf X})) \, A({\bf X}) \;,
\label{a0}
\end{equation}
and define ${\widetilde A}({\bf X}) \equiv A({\bf X}) - A_0$ and
${\widetilde B}({\bf X}) \equiv B({\bf X}) - B_0$.
We then have $C_0(t) = A_0 B_0 + {\widetilde C}_0(t)$, where
\begin{equation}
{\widetilde C}_0(t) \equiv {1\over \rho_0} 
              \int {d^{2f}\!X\over(2\pi\hbar)^f} \,
              \delta(E-H({\bf X})) \, 
              {\widetilde A}({\bf X}(t))
              {\widetilde B}({\bf X})    \;.
\label{ct0}
\end{equation}
Since the system is chaotic (and hence mixing), 
${\widetilde C}_0(t) \to 0$ as $t \to \pm\infty$.
We now have
\begin{equation}
S(E,\Delta) = \rho_0 A_0 B_0 \delta_2(\Delta) 
              + \int_{-\tau_2}^{+\tau_2} {dt\over2\pi\hbar} \,  
                e^{-i\Delta t/\hbar} 
                \left[ \rho_0{\widetilde C}_0(t)
                     + {1\over\pi\hbar} \sum_{\tau_p<\tau_1} 
                       w_p C_p(t) \right].
\label{s6}
\end{equation}
Next, we use the fact that $C_p(t)$ is periodic in $t$ with 
period $\tau_p$, which allows us to write \cite{wilk}
\begin{equation}
C_p(t) = \sum_{k=-\infty}^{+\infty} \Gamma_{pk} \, e^{+2\pi i k t/\tau_p} \;,
\label{cp2}
\end{equation}
where
\begin{equation}
\Gamma_{pk} = {1\over \tau_p} \int_0^{\tau_p} dt \, e^{-2\pi i k t/\tau_p} \, C_p(t) \;.
\label{gpk}
\end{equation}
We note in particular that 
\begin{equation}
\Gamma_{p0} = A_p B_p \;.
\label{gp0}
\end{equation}
Now using Eq.~(\ref{cp2}) in Eq.~(\ref{s6}), we get 
\begin{eqnarray}
S(E,\Delta) &=& \rho_0 A_0 B_0 \delta_2(\Delta) 
              + \rho_0 \int_{-\tau_2}^{+\tau_2} {dt\over2\pi\hbar} \,  
                e^{-i\Delta t/\hbar} \, {\widetilde C}_0(t)
\nonumber \\
            && {} + {1\over\pi\hbar} \sum_{\tau_p<\tau_1} w_p 
                \sum_{k=-\infty}^{+\infty} \Gamma_{pk} \,
                \delta_2(\Delta - 2\pi\hbar k/\tau_p) \;.
\label{s7}
\end{eqnarray}
This result is a slight generalization of Wilkinson's \cite{wilk}.

We now set $\Delta=0$.  For the $k=0$ term in the sum over orbit
modes, we have a factor of $\delta_2(0)=\tau_2/\pi\hbar$; for the 
$k\ne0$ terms, we use Eq.~(\ref{delta2}).  After dividing
through by $\delta_2(0)$, and using $\rho_0=\tau_H/2\pi\hbar$,
the result is
\begin{eqnarray}
{\pi\hbar \over \tau_2} \, S(E,0)
&=& \rho_0 A_0 B_0  
    + {\tau_H\over2\tau_2} \int_{-\tau_2}^{+\tau_2} {dt\over2\pi\hbar}\, 
                                                   {\widetilde C}_0(t)
    + {1\over\pi\hbar}\sum_{\tau_p<\tau_1} w_p\Gamma_{p0} 
\nonumber \\
&& {} + {1\over\pi\hbar} \sum_{\tau_p<\tau_1} {\tau_p\over\tau_2}\,w_p 
   \sum_{k\ne0}{\Gamma_{pk}\sin(2\pi k\tau_2/\tau_p)\over 2\pi k} \;.
\label{s8}
\end{eqnarray}
If we assume $\tau_2\gg\tau_1$, the last term can be neglected.
Comparing Eqs.~(\ref{s4}) and (\ref{s8}),
and recalling Eq.~(\ref{gp0}), we verify Eq.~(\ref{2}).
This is our key result.

There is, however, an important caveat.  We have taken both 
$\tau_1$ and $\tau_2$ to be much greater than 
$\tau_H \sim \hbar^{-(f-1)}$.
The rigorous treatment of \cite{cr,crr}, on the other hand,
requires $\tau_1$ and $\tau_2$ to remain fixed as $\hbar\to 0$,
This suggests that a compromise of $\tau_{1,2}\sim\tau_H$
might be optimal, which is consistent with other analyses of
the trace formula \cite{bk,af,pros,ps}. 
To further investigate this issue,
we consider the special case $A=B$.
The magnitude of fluctuations in the values of the diagonal 
matrix elements of an operator $A$ have been previously 
evaluated \cite{wilk2,em,efkamm,fp,ss2}, with the result
that
\begin{equation}
\rho_0^{(A,A)} = \rho_0 A_0^2 + g\int_{-\tau_H}^{+\tau_H} 
                                {dt\over2\pi\hbar}\;
                                {\widetilde C}_0(t) \;.
\label{p0aa}
\end{equation}
Here $g=2$ if the system is time reversal invariant, and 
$g=1$ if it is not.  Eq.~(\ref{p0aa}) is consistent with
Eqs.~(\ref{s4}) and (\ref{s8}) if we set $\tau_2 = \tau_H/2g$.
We therefore conclude that, in the more general case where
$A\ne B$,
\begin{equation}
\rho_0^{(A,B)} = \rho_0 A_0 B_0 + g\int_{-\tau_H}^{+\tau_H} 
                                   {dt\over2\pi\hbar}\;
                                   {\widetilde C}_0(t) \;.
\label{p0ab}
\end{equation}
If we also set $\tau_1=\tau_2$, then the final term in 
Eq.~(\ref{s8}) will be negligible for short periodic orbits
(although it may become significant as the orbit period 
$\tau_p$ approaches $\tau_2$).
This provides a theoretical explanation for the numerical
results of Main and Wunner \cite{mw}.

Eq.~(\ref{p0ab}) is an explicit formula for the smooth term.
However, for $\tau_{1,2} \sim \tau_H$, it is an uncontrolled
approximation, since the neglected terms in Eqs.~(\ref{s4})
and (\ref{s8}) are not formally suppressed.  Nevertheless,
the agreement of Eq.~(\ref{p0aa})
with the results of \cite{wilk2,em,efkamm,fp,ss2}
leads us to believe that Eq.~(\ref{p0ab}) is correct.

\section{Extensions}
\label{ext}

We now return to Eq.~(\ref{1}) and consider a string of diagonal
matrix elements of $N$ operators, $A,B,\ldots,Z$.
We can find a trace formula for Eq.~(\ref{1}) by starting from
\begin{equation}
F(t_1,\ldots,t_N) = \mathop{\rm Tr}U(t_1)AU(t_2)B \ldots U(t_N)Z \;.
\label{tr2} 
\end{equation}
We then make Fourier transforms with respect to suitable linear
combinations of the $t_i$'s to construct 
\begin{eqnarray}
S(E,\Delta_2,\ldots,\Delta_N) &=&
        \sum_{nml\ldots} A_{nm}B_{ml}\ldots 
        \delta_1(E-{\textstyle{1\over N}}(E_n+E_m+\ldots)) 
\nonumber \\
        && \qquad
           \delta_2(\Delta_2-(E_n-E_m)) \delta_3(\Delta_3-(E_m-E_l)) 
           \ldots \; .
\label{tr3} 
\end{eqnarray}
A straightforward generalization of the analysis in 
Section \ref{analysis} then leads to a result analogous to 
Eqs.~(\ref{s4}) and (\ref{s8}), thus verifying Eq.~(\ref{2}).

The leading contribution to the smooth term is of the form
$\rho_0 A_0 \ldots Z_0$.  There are also subleading contributions 
that depend on various energy-surface correlation functions.  
For example, for $N=3$ we have
\begin{eqnarray}
\rho_0^{(A,B,Z)} &=& \rho_0 A_0 B_0 Z_0 
                   + {g\over2\pi\hbar} \int_{-\tau_H}^{+\tau_H} dt
                     \left[  A_0 {\widetilde C}^{BZ}_0(t)
                           + B_0 {\widetilde C}^{ZA}_0(t)
                           + Z_0 {\widetilde C}^{AB}_0(t) \right]
\nonumber \\
                && {} + {1 \over \rho_0}
                     \left({g\over2\pi\hbar}\right)^2
                     \int_{-\tau_H}^{+\tau_H} dt_1
                     \int_{-\tau_H}^{+\tau_H} dt_2 \,
                                 {\widetilde C}^{ABZ}_0(t_1,t_2) \;.
\label{abz0}
\end{eqnarray}
Here ${\widetilde C}^{AB}_0(t)$ is given by Eq.~(\ref{ct0}),
\begin{equation}
{\widetilde C}^{ABZ}_0(t_1,t_2) = {1\over \rho_0} 
              \int {d^{2f}\!X\over(2\pi\hbar)^f} \,
              \delta(E-H({\bf X})) \, 
              {\widetilde A}({\bf X}(t_1))
              {\widetilde B}({\bf X}(t_2))
              {\widetilde Z}({\bf X})    \;,
\label{abzt0}
\end{equation}
and again $g=2$ if the system is time-reversal invariant, and $g=1$ if it
is not.  This follows from requiring Eq.~(\ref{abz0}) to reproduce 
Eq.~(\ref{p0ab}) when we set $Z=I$.

\section{Conclusions}

We have formulated and verified a precise version of Eq.~(\ref{2}),
which was originally proposed by Main and Wunner \cite{mw}
for chaotic systems.  Our derivation provides an analytic explanation
for their numerical results.

Main and Wunner also proposed an equation analogous to Eq.~(\ref{2})
for integrable systems that generalizes the Berry-Tabor
trace formula \cite{bt}.  We have not attempted to derive
this version, but clearly it would be of interest to do so.

\begin{acknowledgments}
We thank J\"org Main for helpful correspondence.
This work was supported in part by NSF Grant PHY97-22022.
\end{acknowledgments}

\end{document}